\documentclass[]{emulateapj}
\usepackage{natbib}
\usepackage{appendix}
\usepackage{graphicx}
\usepackage{amsmath}
\usepackage[usenames]{color}

\newcommand{\solar}{_{\mathord\odot}}
\newcommand{\hmpc}{\ifmmode{h^{-1}{\rm Mpc}}\;\else${h^{-1}}${\rm Mpc}\fi}
\newcommand{\hMpc}{\ifmmode{h^{-1}{\rm Mpc}}\;\else${h^{-1}}${\rm Mpc}\fi}
\newcommand{\hGpc}{\ifmmode{h^{-1}{\rm Gpc}}\;\else${h^{-1}}${\rm Gpc}\fi}
\newcommand{\hkpc}{\ifmmode{h^{-1}{\rm kpc}}\;\else${h^{-1}}${\rm kpc}\fi}

\newcommand{\msun}{{\rm M}_{\solar}}

\newcommand{\mr}{\ifmmode{M_r}\;\else$M_r$\fi}
\newcommand{\rd}{\ifmmode{R_\delta}\;\else$R_\delta$\fi}
\newcommand{\ngals}{\ifmmode{N_{\rm gals}}\;\else$N_{\rm gals}$\fi}

\begin{document}

\title{{An analytical model for galaxy metallicity: What do metallicity relations tell us about star formation and outflow?}}
\author{Yu Lu\altaffilmark{1}, Guillermo A. Blanc\altaffilmark{1, 2}, Andrew Benson\altaffilmark{1}}
\altaffiltext{1}{The Observatories, The Carnegie Institution for Science,
813 Santa Barbara Street, Pasadena, CA 91101, USA}
\altaffiltext{2}{Departamento de Astronom'a, Universidad de Chile, Camino el Observatorio 1515, Las Condes, Santiago, Chile}

\begin{abstract}
We develop a simple analytical model that tracks galactic metallicities governed by star formation and feedback to
gain insight from the observed galaxy stellar mass-metallicity relations over a large range of stellar masses and redshifts.  
The model reveals the following implications of star formation and feedback processes in galaxy formation.
First, the observed metallicity relations provide a
stringent upper limit for the averaged outflow mass-loading factors of local galaxies, which is $\sim20$ for $M_*\sim10^9\msun$
galaxies and monotonically decreases to $\sim1$ for $M_*\sim10^{11}\msun$ galaxies.  Second, the inferred upper-limit for the
outflow mass-loading factor sensitively depends on whether the outflow is metal-enriched with respect to the ISM metallicity.  If
half of the metals ejected from SNe leave the galaxy in metal-enriched winds, the outflow mass-loading factor for galaxies at any
mass can barely be higher than $\sim10$, which puts strong constraints on galaxy formation models.  
Third, the relatively lower stellar-phase to gas-phase metallicity ratio for lower-mass galaxies 
indicate that low-mass galaxies are still rapidly enriching their metallicities in recent times,  while high-mass galaxies are more 
settled, which seems to show a downsizing effect in the metallicity evolution of galaxies. 
The analysis presented in the paper demonstrates the importance of accurate
measurements of galaxy metallicities and the cold gas fraction of galaxies at different redshifts for constraining star formation
and feedback processes, and demonstrates the power of these relations in constraining the physics of galaxy formation.
\end{abstract}

\section{Introduction}\label{sec:introduction}

Star formation and galactic outflows triggered by feedback from star formation
are considered the most important processes in galaxy formation \citep{Benson2010c, Houjun-Mo2010a}. 
Galaxy surveys find that galaxies comprise only a small fraction of baryonic matter in the Universe 
\citep[e.g.][]{Brinchmann2000a, Cole2001a, Dickinson2003a, Bell2003a}. 
Assuming galaxies form in Cold Dark Matter (CDM) halos \citep{Blumenthal1984a}, 
the baryon mass fraction in halos hosting a galaxy smaller than the Milky Way must 
decrease rapidly for decreasing halo mass in order to explain the shallow slope of the low-mass end 
of galaxy mass function, pointed out in pioneering works by \citet[][]{Frenk1988a, Cole1991a, White1991a}, 
and elaborated by recent analysis \citep[e.g.][]{Papastergis2012a, Behroozi2013a, Moster2013a, Lu2015c}. 
Strong outflows have been proposed as the most important process responsible for keeping low-mass 
galaxies baryon poor \citep[e.g.]{Dekel1986a, Lacey1991a}. 
Many galaxy formation models successfully reproduce the number density of low-mass 
galaxies by invoking strong outflows \citep[e.g.][]{Benson2003a, Somerville2008a, Guo2011c}. 
Using model inference techniques, \citet{Lu2014a} and \citet{Benson2014a} 
found that to match the data the outflow rate needs to be more than 10 times higher than the star formation rate 
for low-mass galaxies. 
Moreover, these outflows must be sustained, or at least be recurrent, over cosmological time scales. 
Although strong feedback in starburst galaxies is often observed, it is not clear if the duty cycle of the outflow 
is sufficiently high to be consistent with what seems to be required in the models.
In spite of the success of the assumption of strong outflows in reproducing many statistical properties of the galaxy 
population, observational evidence for such strong outflows in local galaxies is still lacking \citep[e.g.][]{Bouche2012a, Kacprzak2014a}.
Therefore, it is important to seek other independent observational tests to constrain the strength of outflows. 

Metals other than the primordial species can only be produced by star formation. Therefore, metallicities of the baryonic 
matter in different phases in a galaxy are expected to put interesting constraints on star formation and outflow. 
For a given galaxy, the total stellar mass provides the total metal mass budget to be distributed anywhere associated with 
the galaxy (including in outflows). The total amount of metal mass remaining in a galaxy, therefore, contains critical information 
about how outflows work. Moreover, how metals are partitioned between different phases of baryonic matter may give some 
insight into the nature of star formation. Using the combination of gas- and stellar-phase metallicities, one can attempt to 
infer the broad nature of the star formation history in a galaxy.

To extract these pieces of information from observational data, we develop an analytic model that can logically 
connect star formation and outflow processes with the observed metallicity relations. 
Using this model, we attempt to gain insight into these metallicity relations 
and to constrain the strength of outflow and the metallicity of past star formation. 
We find that with simple but plausible assumptions 
such a model can match the observed metallicity relations and draw inferences about star formation and outflow 
based on existing data.

The paper is organized as follows. 
In \S\,\ref{sec:method}, we describe the model of the evolution of metal content of galaxies which is
based on minimal assumptions. We develop two different approaches to constrain the mass-loading factors
of galactic outflows and star formation in different environments using data on the mass-metallicity relations. 
In \S\,\ref{sec:data}, we describe the observational data adopted in this paper. 
We then demonstrate in \S\,\ref{sec:result} how observational data on the stellar- and gas-phase metallicities
constrain the outflow mass-loading factor and star formation. 
In \S\,\ref{sec:discussion}, we discuss the conclusions and implications of these results. 
Throughout the paper, we assume Solar metallicity $Z_{\odot}=0.0134$ \citep{Asplund2009a} and Solar oxygen abundance  
$12+\log(O/H)_{\odot}=8.69$ \citep{Allende-Prieto2001a, Asplund2009a}. 
We adopt a Chabrier IMF \citep{Chabrier2003a} for all data and modeling. 
We have converted all the adopted data based on these assumptions. 
Another important quantity we need to fix is the chemical yield, $y$, which is defined as the ratio between the mass of newly 
produced metals that are ejected into the interstellar gas and the mass locked in long-lived stars for a single stellar population, 
as originally defined by \citet{Searle1972a}. 
Note that this is different from another widely adopted definition of yield in the literature, $p$, which is defined as the mass of newly 
produced metals per unit gas mass that is turned into stars (i.e. these two definitions differ due to the recycling of gas from short-lived stars). 
In this paper, we choose to use a rather high metal yield, $y=0.07$, which corresponds to $p=0.038$ for the Chabrier IMF. 
This value is at the higher end of values predicted by nucleo-synthesis models \citep{Woosley1995a, Woosley2002a, Nomoto1997a, Nomoto1997b}.
We choose to use this high yield to have a conservative estimate of the upper limit of the inferred outflow mass-loading factor.

\section{An analytic model for galaxy metallicity}\label{sec:method}
The origin and implication of galaxy metallicity relations are extensively studied by many authors
using analytical models \citep[e.g.][]{Tinsley1978a, Matteucci1987a, Koppen1999a, Dalcanton2007a, 
Erb2008a,  Dave2012a, Peeples2013a,  Pipino2014a, Zahid2014a}, 
semi-analytic galaxy formation models  \citep[e.g.][]{Cole2000a, De-Lucia2004a, Nagashima2005a, Yates2013a}, 
and hydrodynamical simulations \citep[e.g.][]{Finlator2008a, Dave2011b, Kobayashi2007a, Wiersma2009a, Ma2015a}. 
In this paper, we take the analytical approach to build a simple model that connects the buildup of metal mass 
in a galaxy and outflows of material from that same galaxy. 
In this model, star formation happens in the interstellar medium (ISM), which
is continuously enriched with metals by star formation. 
We assume that metals are instantaneously produced by star formation, 
and the time delays between the nuclear synthesis of different heavy elements are ignored for simplicity
(i.e. the usual ``instantaneous recycling approximation''). In addition, we assume that the mixing of metal 
mass in the ISM is perfect. For simplicity, we further assume that the newly accreted gas 
does not contribute any significant metal mass. 

Under these assumptions, we can write down a set of equations to follow the metal mass. 
The change of the metal mass in the ISM when ${\rm d}M_*$ of stellar mass is 
formed can be written as
\begin{equation}\label{equ:model_basic}
{\rm d} M_{\rm z,g} = y {\rm d} M_*  - Z_{\rm g} {\rm d} M_* - {\eta \over 1-R} Z_{\rm g} {\rm d}M_*\,,
\end{equation}
where $M_{\rm z,g}$ denotes the metal mass in cold gas, $M_*$ denotes the mass of long-lived stars, 
$y$ is the chemical yield, defined as the ratio between the mass of newly 
produced metals and the mass locked in long-lived stars \citep{Searle1972a}, 
$Z_{\rm g}$ is the gas-phase metallicity at the time of star formation,  
$R$ is the mass fraction that is returned into ISM from short lived stars and stellar wind, 
and $\eta$ is the galactic outflow mass-loading factor, which is defined as 
the ratio between the outflow mass flux and the instantaneous star formation rate. 
The first term in the right side of the equation represents the metal mass newly formed and returned 
by star formation; the second term represents the metal mass locked into long-lived stars; 
the third term represents the metal mass in the ISM that is carried away by outflows. 

We note that the chemical yield $y$ and the return fraction $R$ are approximately constant parameters, 
which are largely determined by the stellar initial mass function (IMF) and weakly depend on metallicity \citep[e.g. BC03][]{Bruzual2003a}. 
For a Chabrier IMF \citep{Chabrier2003a}, $R=0.46$ is relevant to adopt for the instantaneous recycling approximation. 
The outflow mass-loading factor $\eta$ is a variable to be constrained in this paper. 
In principle, the outflow mass-loading factor $\eta$ is a variable that can change 
with time and from galaxy to galaxy. In this paper, 
we assume that $\eta$ can arbitrarily vary as a function of the present-day galaxy stellar mass, 
but is a constant for a given galaxy (i.e. is not a function of time). 
This is equivalent to assuming that the parameter $\eta$ is an effective mass-loading factor 
averaged over the history of the galaxy and weighted by the star formation rate, i.e.
$\eta = {\int_0^{t_0} \eta(t) \phi(t) {\rm d}t / \int_0^{t_0} \phi(t) {\rm d}t}$, where $\phi(t)$ 
is the time dependent star formation rate. 
We note that because stellar mass can only increase for an isolated galaxy, 
one can use stellar mass as a clock to integrate the increase of the metal mass over time. 

Using this simple model, we can derive a set of equations to describe the relationship 
between metallicities and parameters characterizing star formation and outflow. 
In the first set of derivations, we seek the relationship between the stellar-phase metallicity, 
gas-phase metallicity and the outflow mass-loading factor. 
The change of the metal mass locked into long-lived stars in a time interval is the stellar mass 
formed in the time interval multiplied by the instantaneous gas-phase metallicity $Z_{\rm g}$, as 
\begin{equation}\label{equ:model_mzstar}
{\rm d} M_{\rm z,*} = Z_{\rm g} {\rm d}M_*\,.
\end{equation}
From this equation, one realizes that the term 
$Z_{\rm g}{\rm d}M_*$ in Equation (\ref{equ:model_basic}) is just the metal mass 
that is locked into long-lived stars. 
The integral of Equation (\ref{equ:model_basic}) immediately yields that
\begin{equation}
M_{\rm z, g}  = y M_* - \left(1+{\eta \over 1-R}\right) Z_* M_*\,,
\end{equation}
where $Z_*$ is the averaged stellar-phase metallicity. 
Adopting the definition for the gas-phase metallicity, $Z_{\rm g} = {M_{\rm z,g} / M_{\rm g}}$, 
we find that the gas-phase metallicity is
\begin{equation}
Z_{\rm g} = {y \over r_{\rm g}} - \left(1+{\eta \over 1-R} \right) {Z_* \over r_{\rm g}}\,,
\end{equation}
where $r_{\rm g}$ is the gas-to-stellar mass ratio of galaxies at the present day, e.g. $r_{\rm g} = M_{\rm g} / M_*$.
We note that how the gas-to-stellar mass ratio evolves in the past is irrelevant here, because this equation 
only describes the present-day metallicities of a galaxy.
This equation shows that the gas-phase metallicity decreases with an increasing gas-to-stellar mass ratio, 
as the cold gas dilutes the metallicity of the ISM. 
It also shows that the gas-phase metallicity decreases with an increasing outflow mass-loading factor, 
as outflow expels metal-enriched gas out of galaxies. 
We can rewrite the equation as follows to determine the mass-loading factor using metallicities, 
\begin{equation}\label{equ:eta_app1}
\eta = (1-R) \left[ {y \over Z_*} - {Z_{\rm g} \over Z_*} r_{\rm g} - 1\right]\,.
\end{equation}
This equation shows that, for a given galaxy, higher metallicities and higher cold gas-stellar mass ratio 
indicate a lower outflow mass-loading factor. This is easy to understand because strong outflow will reduce 
the metal content and gas content of the galaxy.

In the second set of derivations, we adopt the gas-phase metallicity at different redshifts as data constraints.  
To do so, we need to assume how the cold gas metallicity $Z_{\rm g}$ evolves with stellar mass growth. 
Here we first adopt a simple assumption that the gas-phase metallicity follows a power-law trajectory in the $Z_{\rm g}$--$M_*$ plane 
as the galaxy evolves, e.g. 
\begin{equation}\label{equ:zgas_powerlaw}
Z_{\rm g}(t) = Z_{\rm g} \left({M_*(t) \over M_*}\right)^\mu \,,
\end{equation}
in which the power index, $\mu$, is assumed to vary as a function of galaxy stellar mass, and is to be determined. 
The variation of $\mu$ reflects different ways in which galaxies can evolve their metallicities as their stellar masses 
increase. For example, inflow can decrease metallicity, and outflow can increase or decrease the metallicity depending on 
how metals are loaded in the wind. 
The assumption of power-law trajectories allows galaxies with different masses to evolve along different paths. 
Based on this simple assumption, by integrating Eq.\ref{equ:model_basic} we find that
\begin{equation}\label{equ:mzgas}
\begin{split}
M_{\rm z,g} = y M_* - \left(1+{\eta \over 1-R}\right) {1 \over \mu+1} Z_{\rm g}M_*\,.
\end{split}
\end{equation}
Applying the definition of the gas-phase metallicity,
we can rewrite the equation to be an expression for the mass-loading factor, 
\begin{equation}\label{equ:eta_app2}
\eta = (1-R)\left[\left(\mu+1\right) \left({y \over Z_{\rm g}} - r_{\rm g}\right) - 1\right]\\.
\end{equation}
This equation shows that we can also constrain the averaged mass-loading factor $\eta$, 
when the trajectory of a galaxy in the $Z_{\rm g}-M_*$ plane is determined. 

In the above derivations, we have assumed that all the newly produced metals are returned into the ISM, so that the yield 
affecting the chemical evolution of a galaxy equals to the intrinsic nucleo-synthesis yield. Because SN ejecta can preferentially 
transport over-enriched (relative to the mean ISM) gas out of the potential well of a galaxy as shown in hydro-dynamical simulations 
\citep[e.g.][]{Creasey2015a, Melioli2015a}, it is reasonable to assume that the ``retained'' yield, $\tilde{y}$, is less than the intrinsic yield, 
$y$. In the following analysis, we vary the retained yield to explore this effect. 

\section{Data}\label{sec:data}
\begin{figure}[htb]
\begin{center}
\includegraphics[width=0.45\textwidth]{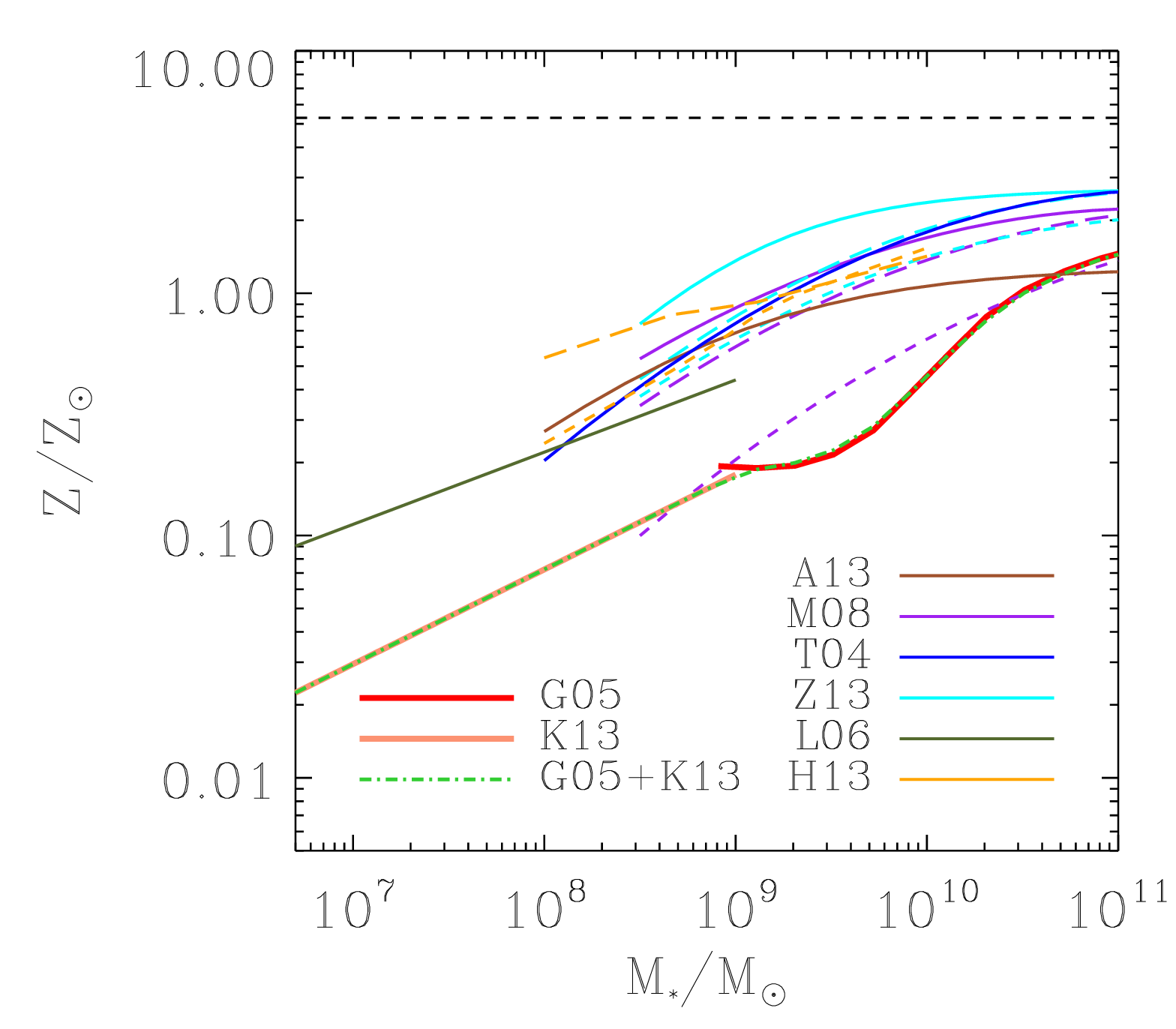}
\caption{Observational results of metallicity as a function of galaxy stellar mass. The stellar-phase metallicity relations are from \citet{Gallazzi2005a} and \citet{Kirby2013a}, 
denoted by the solid lines of G05 and K13, respectively. 
The green dash-dotted line shows a smooth function joining both of the data sets. 
The local galaxy gas-phase metallicity relations are from \citet{Tremonti2004a}, \citet{Maiolino2008a}, \citet{Zahid2013a}, \citet{Andrews2013a}, and \citet{Lee2006a}, 
which are denoted by colored solid lines of T04, M08, Z13, A13, and L06, respectively. 
The gas-phase metallicity relations at higher redshifts of \citet{Maiolino2008a} and \citet{Zahid2013a} are shown by long-dashed lines for $z\sim 0.7$ 
and short-dashed lines for $z\sim2.2$. 
The gas-phase metallicity relations of \citet{Henry2013a} for $z\sim 0.6$ and \citet{Henry2013b} for $z\sim1.8$, denoted by H13, are shown 
by long-dashed and short-dashed orange lines. 
The horizontal short dashed line denotes the intrinsic yield $y$ assumed in the paper. 
}
\label{fig:data}
\end{center}
\end{figure}

The derivations presented in the last section demonstrate that one can draw inferences about star formation and feedback 
when certain observational data are given. Based on our derivations, these observational data include the stellar-phase 
metallicity as a function of galaxy stellar mass, the gas-phase metallicity as a function of galaxy stellar mass of local galaxies, 
and of galaxies at different redshifts. To demonstrate the constraining power of these observational relations based on 
the analytic derivations shown in the previous section, we choose to use the mean relation of each of observational results. 
For the stellar-phase metallicity, we adopt the observational results of \citet{Gallazzi2005a} and \citet{Kirby2013a} 
to cover a wide range of stellar masses between $\sim10^7\msun$ and $10^{11}\msun$. 
As shown by \citet{Kirby2013a}, these two pieces of data join with each other remarkably well if 
the relation determined by the Kirby data is extrapolated to high stellar masses. 
In the analysis of this paper, we adopt a function that follows the two data sets and smoothly 
joins them as shown in Fig. \ref{fig:data}. 
We note that because the uncertainty in the stellar metallicities at the low mass end of 
the \citet{Gallazzi2005a} relation is large, the small deviation between their mean relation 
and our adopted function is ignored in this paper. 
In addition, there are two biases in the stellar-phase metallicity. 
First, the observed metallicity is weighted by young and luminous stars, which tend to be metal rich. 
Second, the stellar-phase metallicity measurements are mainly sensitive to iron abundances rather than the $\alpha$ elements, 
which are the species measured in the gas-phase metallicity measurements.  
Following \citet{Peeples2014a}, we have corrected these biases  
using their Eq.(7) and Eq.(A1), taking into account the different solar metallicity adopted in the reference paper.
For the gas-phase metallicity relations, there are a larger number of results published in the literature. 
While the observed gas-phase metallicity relations have small scatter at a given redshift (typically 0.1~dex), 
they have substantial differences in their normalizations and shapes owing to large uncertainties in 
the calibration of the metallicity measurements \citep{Kewley2008a}. 
To take these uncertainties into account in our inferences, we adopt multiple observational results. 
For the gas-phase metallicity--stellar mass relation of local star forming galaxies, 
we adopt the results of \citet{Tremonti2004a, Maiolino2008a, Zahid2013a} and \citet{Andrews2013a}
for galaxies with stellar mass in a range between $\sim10^9\msun$ and $10^{11}\msun$, 
and \citet{Lee2006a} for lower mass galaxies in the range between $\sim 10^7\msun$ and $10^9\msun$. 
For higher redshifts, we adopt the metallicity-stellar mass relation at $z\sim2.2$ compiled by \citet{Maiolino2008a} 
and \citet{Zahid2013a}, respectively. In addition, we also use the gas-phase metallicity 
measurements of \citet{Henry2013a, Henry2013b} for a lower mass range between $10^8\msun$ and $10^{10}\msun$
as a complement. To infer the mass-loading factor with the approaches demonstrated in the previous section, 
one also needs to know the gas-to-stellar mass ratio of galaxies. We adopt a fitting formula for the mean relation 
of star forming galaxies compiled by \citet{Peeples2014a}, namely
\begin{equation}
\log{r_{\rm g}} = -0.48 \log \left({M_* \over \msun}\right) + 4.39\,. 
\end{equation}
As shown by the authors, this fitting formula captures the mean relation between the gas mass ratio 
and galaxy stellar mass well and agrees very well with many observational estimates \citep[e.g.][]{McGaugh2005a, McGaugh2012a, Leroy2008a, Papastergis2012a}.  
Unlike the metallicity-stellar mass relations, the cold-gas to stellar mass ratio has a significantly large scatter, 
$\sim0.5$dex. We add this scatter into the gas mass ratio as upper and lower bounds with $\Delta\log{r_{\rm g}}=0.5$
to demonstrate how this scatter propagates into our results. 
For all the data sets, we have corrected the stellar mass by assuming a \citet{Chabrier2003a} IMF.

\section{Results}\label{sec:result}

\subsection{Outflow mass-loading factor Inferred from local metallicity relations}\label{sec:result_app1}

\begin{figure}[htb]
\begin{center}
\includegraphics[width=0.45\textwidth]{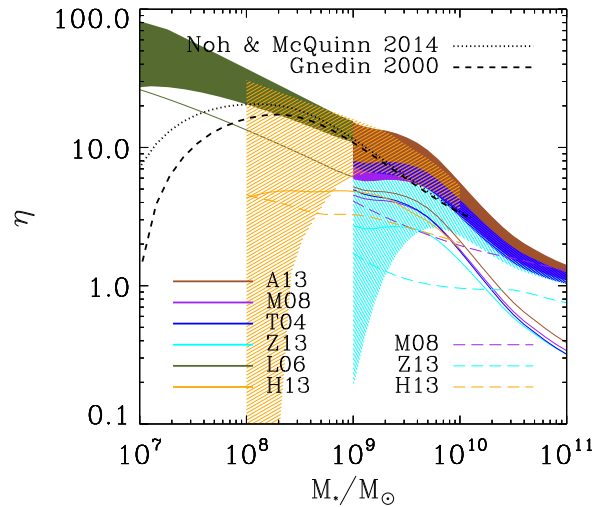}
\caption{The constrained outflow mass-loading factor $\eta$ derived from different combinations of observational data.
The shaded regions and the solid lines are the results inferred from the stellar-phase metallicity and the gas-phase metallicity 
of local galaxies. 
The shaded regions show $\eta$ obtained by assuming that all produced metal mass is mixed 
into the ISM but varying the cold gas to stellar mass ratio. 
The upper and lower bounds encompass the variations resulted from the $\pm0.5$dex 
scatter of the cold gas mass ratio for given stellar mass. 
The solid lines show the results assuming half of the produced metal is directly expelled from galaxies, i.e. $\tilde{y}=0.5y$ 
with the fiducial cold gas mass ratio. 
The dashed lines denote the results inferred from the gas-phase metallicity relation at multiple redshifts (see \S\ref{sec:result_app2}).  
Different colors denote different data 
sources. The brown line represents the result using \citet{Andrews2013a} data. 
The purple lines denote the results using \citet{Maiolino2008a} data. 
The blue line shows the result using \citet{Tremonti2004a} data. 
The cyan lines show the results using \citet{Zahid2013a} data.
The orange line shows the results using \citet{Henry2013a, Henry2013b} data. 
The green line shows the result using \citet{Lee2006a} data.
The black dotted and dashed lines denote the upper limit estimates of the mass-loading factor allowed by 
two different re-ionization models. }
\label{fig:eta}
\end{center}
\end{figure}

As we demonstrated in \S\ref{sec:method}, without assuming any particular star formation history or 
metallicity enrichment history, our simple model can match the stellar-phase and gas-phase metallicity 
relations at a given redshift and can constrain the mass-loading factor. 
In this subsection, we adopt the combined the stellar-phase metallicity as a function of stellar mass and 
the gas-phase metallicity results of \citet{Tremonti2004a, Maiolino2008a, Zahid2013a, Andrews2013a}, 
and \citet{Lee2006a} of local galaxies to infer the outflow mass-loading factor, $\eta$, as a function of galaxy stellar mass.  
We also include the gas-phase metallicity relation of \citet{Henry2013a} at $z=0.6$ in this analysis. 
We choose each of the gas-phase metallicity relation and combine it with the stellar-phase metallicity 
relation and the gas-to-stellar mass ratio to feed into Equation (\ref{equ:eta_app1}). 
At first, we assume all of the metals produced are retained in the galaxy and mixed into ISM 
before they can be expelled by outflows, i.e $\tilde{y}=y$. 
In other words, outflows have the same metallicity as the ISM with no further metallicity enhancement.
We vary the normalization of the gas-to-stellar mass ratio, $r_{\rm g}$, up and down by 0.5~dex to 
encompass the random variation of the gas mass for given stellar mass. 
The resulting mass-loading factor as a function of galaxy stellar mass for each data set is shown by 
the shaded bands in Fig. \ref{fig:eta}. 
For the all results, the general trend is that the mass-loading factor is higher for 
low-mass galaxies and decreases rapidly for higher stellar masses. As one can see, the inferred 
$\eta$--$M_*$ relations by using different gas-phase metallicity relations are very similar to each other 
at the high-mass end, where the inferred mass-loading factor is less than 2. 
Another interesting feature is that the inferred mass-loading factor tends to flatten out at the very high mass end. 
The reason for this is that mergers become increasingly important for higher mass galaxies, and 
the mass-loading factor we constrain is considered to be an average over all progenitors, which have relatively lower masses. 
In the lower mass regime, 
the scatter in the gas mass ratio produces increasingly larger variations in the inferred outflow 
mass-loading factor. Nevertheless, the upper bound for $\eta$ is well defined. For galaxies 
$M_*=10^9\msun$, the upper-bound mass-loading factor inferred from all the data combinations is 
less than 20. For the same stellar mass, galaxies with higher gas fraction may have significantly lower $\eta$. 
In addition, different gas-phase metallicity results have different level of sensitivity to the variation of 
the gas mass ratio. A higher gas-phase metallicity not only suggests a systematically lower 
mass-loading factor, but also a more sensitive dependence on the gas mass ratio. 
Second, we assume that SN ejecta carry away half of the produced metal mass without mixing it into ISM. 
In this situation, $\tilde{y}=0.5y$, and we show the inferred $\eta$ with the median gas mass ratio 
by solid lines in Fig.\ref{fig:eta}. When a fraction of metal leaks from galaxies without mixing into ISM, 
the observed metallicity relations yield a much lower $\eta$. At the high-mass end, all the data 
combinations suggest that the mass-loading factor is below 1. At $M_*=10^9\msun$, the mass-loading factor 
is below 10, and it increases with decreasing stellar mass. At the low-mass end of the mass range 
($\sim10^7\msun$), the mass-loading factor is about 20. It is worth noting that we chose to use a rather 
high metal yield, $y=0.07$, which is about five times of the Solar metallicity, 
and at the high end of usual values predicted by neucleo-synthesis models \citep{Woosley1995a, Nomoto1997a, Nomoto1997b}. 
We chose this high yield parameter to demonstrate an upper limit for the mass-loading factor. 
If a lower metal yield is chosen instead, the inferred mass-loading factor decreases.

For $10^7\msun$ galaxies, the mass-loading factor inferred by the \citet{Lee2006a} gas-phase metallicity 
and the \citet{Kirby2013a} stellar-phase metallicity relations can be as high as $\sim20$. 
We note, however, that low-mass galaxies are expected to be 
hosted by low-mass halos, and re-ionization  
would prevent a large fraction of baryonic mass from collapsing into such low-mass halos in the first place, 
so that galaxies forming in those halos would never have contained as much baryonic mass as these large 
mass-loading factors implied, unless the outflow materials are rapidly reaccreted back into the galaxy. 
We note that $\eta$ is the averaged net mass-loading factor, while the instantaneous mass loading 
in individual starburst galaxies may be higher if there is significant baryon mass recycled from early times.  
Under the assumption that reaccretion is not important, the total baryonic mass in stars, cold gas, 
and ejected by outflow should be lower than the total baryonic mass that can collapse into the host halo, i.e.
\begin{equation}
M_* + M_* r_{\rm g} + \eta {M_* \over 1-R} \leq f_{\rm b} f_{\rm reion}(M_{\rm vir}, z) M_{\rm vir}\,,
\end{equation}
where $f_{\rm b}=0.17$ is the cosmic baryon fraction, and $f_{\rm reion}$ is the fraction of baryon mass 
that can collapse into a halo with virial mass $M_{\rm vir}$ at redshift $z$ due to re-ionization. 
This inequality yields that
\begin{equation}\label{equ:reionization}
\eta \leq (1-R) \left[ f_{\rm b} f_{\rm reion} {M_{\rm vir} \over M_*} - r_{\rm g} - 1\right]. 
\end{equation}
To estimate the limit for the mass-loading factor, we adopt a re-ionization model by \citet{Gnedin2000a} 
with a fitting formula proposed by \citet{Kravtsov2004a} and a recent model by \citet{Noh2014a} 
to compute the total baryon mass fraction as a function 
of halo mass at $z=0$. We then adopt the abundance matching model of \citet{Behroozi2013a} and extrapolate 
it to the low-mass end ($M_*\sim 10^7\msun$) to determine the stellar mass--halo mass relation. 
Using these and Equation (\ref{equ:reionization}), we can derive an upper limit for the mass-loading factor allowed 
by re-ionization. The thick dashed line and dotted line in Fig.\ref{fig:eta} show the upper limits set by the Gnedin model 
and the Noh \& McQuinn model, respectively. 
In both cases, the limiting outflow mass-loading factor drops sharply when galaxy stellar mass goes below $\sim10^8\msun$. 
Below this mass scale, re-ionization prevents baryons from collapsing into the low-mass halos hosting these galaxies. 
These low-mass galaxies should therefore never have contained as many baryons as is implied by their inferred 
mass-loading factors given their existing stellar mass and cold gas mass. 
To reconcile the observed metallicities and cold baryon masses of low-mass galaxies, a more plausible model 
seems to require metal-enriched outflows, which carry away a large fraction of metals directly from SN ejecta 
without mixing much metals win ISM. In this scenario, the retained metal yield can be lower than we used here, 
yielding lower mass-loading factors.

\begin{figure}[htbp]
\begin{center}
\includegraphics[width=0.45\textwidth]{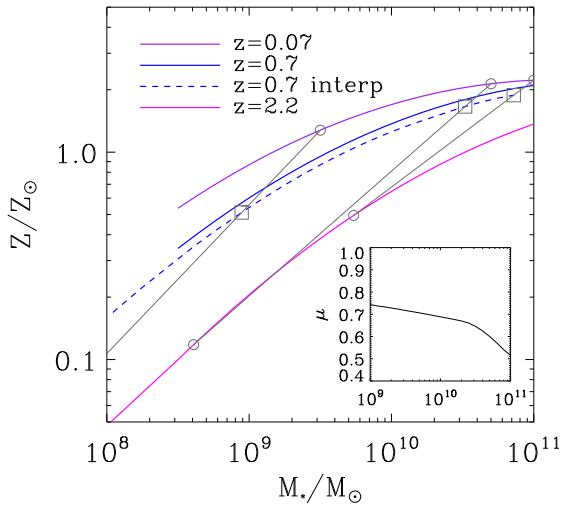}
\caption{ The grey lines show three examples of the power-law trajectories of galaxies in the $Z_{\rm g}$--$M_*$ diagram. 
These trajectories are determined by assuming star formation histories parameterized by Equation (\ref{equ:sfr}) and 
the gas-phase metallicity relations at $z=0.07$ and $z=2.2$ from \citet{Maiolino2008a}. 
The blue dashed line shows the predicted gas-phase metallicity as a function of stellar mass at $z=0.7$ using the determined trajectories. 
The solid lines are the observational results of \citet{Maiolino2008a} at different redshifts as noted in the legend.
The inserted diagram shows the inferred $\mu$ (the power index for the metallicity--stellar mass trajectory) as a function of final stellar mass 
in units of $\msun$. 
}
\label{fig:ztrack}
\end{center}
\end{figure}

\subsection{Outflow mass-loading factor inferred from gas-phase metallicity relations at different redshifts}\label{sec:result_app2}

We now use the gas-phase metallicity relation at multiple redshifts to carry out the inference 
described in \S\ref{sec:method}. We adopt the gas-phase metallicity--stellar mass relation 
compiled by \citet{Maiolino2008a} for galaxies at two different redshifts $z=0.07$ and $z=2.2$, 
the relation compiled by \citet{Zahid2013a} at $z=0.08$ and $z=2.3$, 
and the relation measured by \citet{Henry2013a, Henry2013b} at $z=0.6-0.7$ and $z=1.3-2.3$ 
for lower stellar masses ($M_*=10^8-10^{10}\msun$). 
As we described in \S\ref{sec:method}, we use this data to determine 
the trajectories along which galaxies evolve in the $Z$--$M_*$ diagram.
For the simple case we demonstrated in \S\ref{sec:method}, we essentially need to 
determine the logarithmic slope, $\mu$, as a function of galaxy stellar mass 
using the observed gas-phase metallicity relations at two separate redshifts. 
One way to determine $\mu(M_*)$ is to use a realistic star formation history of 
a galaxy with a stellar mass $M_*$ at a lower redshift to determine the stellar mass of 
its typical progenitor at a higher redshift, and then determine where the progenitor galaxy 
is in the $Z_{\rm g}$--$M_*$ diagram at the higher redshift by using the gas-phase 
metallicity--stellar mass ration at that redshift. 
We adopt a fitting model for the star formation histories of star forming galaxies proposed by \citet{Leitner2012a}, 
\begin{equation}\label{equ:sfr}
\phi(M_*, z) = A_0 \left({M_* \over 10^{11} \msun} \right)^{\beta+1} (1+z)^{\alpha}\,, 
\end{equation}
where parameters $A_0=3.24\msun{\rm yr}^{-1}$, $\alpha=3.45$, and $\beta=-0.35$ are determined 
by matching the abundances of star forming galaxies out to high redshift ($z\sim4$).
Using this model, we can integrate the star formation rate history to predict the stellar mass at any 
given redshift for a galaxy with a given final stellar mass. For simplicity, we again assume instantaneous 
recycling when integrating SFR histories. For more accurate models taking into account time-dependent 
recycling, readers are referred to \citet{Leitner2011a}, \citet{Leitner2012a} and \citet{Lu2015a}. 
Assuming a galaxy evolves along a power-law trajectory on the gas-phase metallicity-stellar mass plain between 
two different redshifts, we can find where the trajectory determined by the star formation history intersects 
with the gas-phase metallicity-stellar mass relations at the different redshifts. Using the two intersection points in 
the $Z_{\rm g}-M_*$ diagram, we can determine the logarithmic slope as 
$\mu=(\log{Z_{\rm g}^0} - \log{Z_{\rm g}^1}) / (\log{M_*^0}-\log{M_*^1})$, 
where the superscripts ``0'' and ``1'' denote quantities at two different redshifts of a galaxy. 
Fig.\ref{fig:ztrack} shows an example of how the $\mu$ parameter is determined as a function of stellar mass 
using the \citet{Maiolino2008a} data. 
In the figure, we show three trajectories with different final galaxy masses in grey lines and 
over-plot them with the gas-phase metallicity relation of \citet{Maiolino2008a} at $z=0.07$, 0.7, and 2.2. 
For each given final stellar mass at $z=0.07$, we compute the stellar mass at $z=2.2$ using the star 
formation history of Eq.\ref{equ:sfr}. We read the gas-phase metallicity of the galaxy with the stellar mass 
from the gas-phase metallicity relation at the corresponding redshift. The open circles along a grey line are 
the positions of the galaxy at the two redshifts, $z=0.07$ and 2.2. 
The grey lines connecting the two circles are the power-law trajectories of the three example 
galaxy masses. The grey squares between the two circles on each trajectory mark the predicted gas-phase
metallicity--stellar mass relation at $z=0.7$. We predict those quantities at the redshift by interpolating along 
the star formation history and the power-law $Z$--$M_*$ trajectory. 
Connecting these predictions, the dashed line shows the predicted gas-phase metallicity--stellar mass relation at $z=0.7$. 
As one can see, the interpreted gas-phase metallicity relation at $z=0.7$ agrees with the observational result at 
the same redshift remarkably well, which demonstrates that the power-law trajectory sufficiently well captures 
the evolution of a galaxy in the $Z_{\rm g}$--$M_*$ plane. 
In the inserted panel in Fig.\ref{fig:ztrack}, we show the determined $\mu$ as a function of final galaxy stellar mass. 
The resulting $\mu$ slowly decreases from 0.73 at $M_*=10^9\msun$ to 0.51 at $M_*=10^{11}\msun$. 
\citet{Peeples2013a} have performed more detailed inference on the gas-phase metallicity trajectories 
by assuming a fundamental metallicity--stellar mass--SFR 
relation \citep{Mannucci2010a} without enforcing the trajectories to follow a power-law form. 
From Figure 1 of \citet{Peeples2013a}, one can find that their trajectories are nearly power-law over a large range of redshifts, 
which in turn supports our simple assumption of power-law trajectories made in Equation (\ref{equ:zgas_powerlaw}).

Combining the determined $\mu(M_*)$ and the data and inserting these terms into Equation (\ref{equ:eta_app2}),
we can again compute the mass-loading factor as a function of stellar mass. 
The results are shown as long dashed lines in Fig.\ref{fig:eta}. 
As one can see, the results obtained from this approach are qualitatively similar to those inferred without assuming 
a star formation history in \S\ref{sec:result_app1}.
The constrained mass-loading factor decreases with increasing stellar mass. 
Quantitatively, using the same yield, $\tilde{y}=y=0.07$, the mass-loading loading factors 
inferred in this approach are a factor of 2-3 lower than those inferred in 
\S\ref{sec:result_app1}. 
The difference may be due to the calibration of different data sets or inconsistency between the 
empirical star formation history and the real star formation history of the star-forming galaxies included 
in the gas-phase metallicity measurements.

\subsection{Metallicity history}

\begin{figure*}[htbp]
\begin{center}
\includegraphics[width=0.42\textwidth]{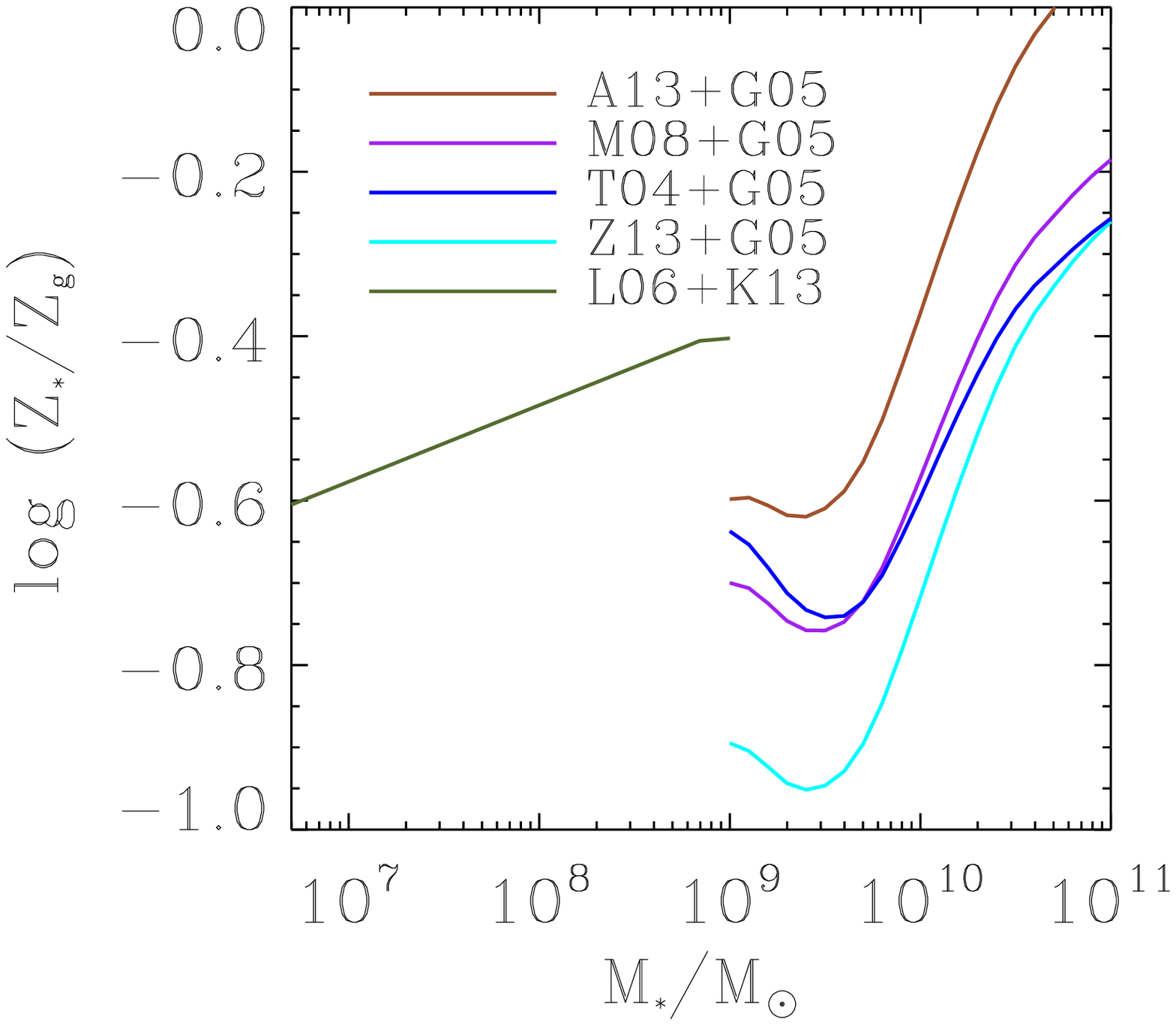}
\includegraphics[width=0.45\textwidth]{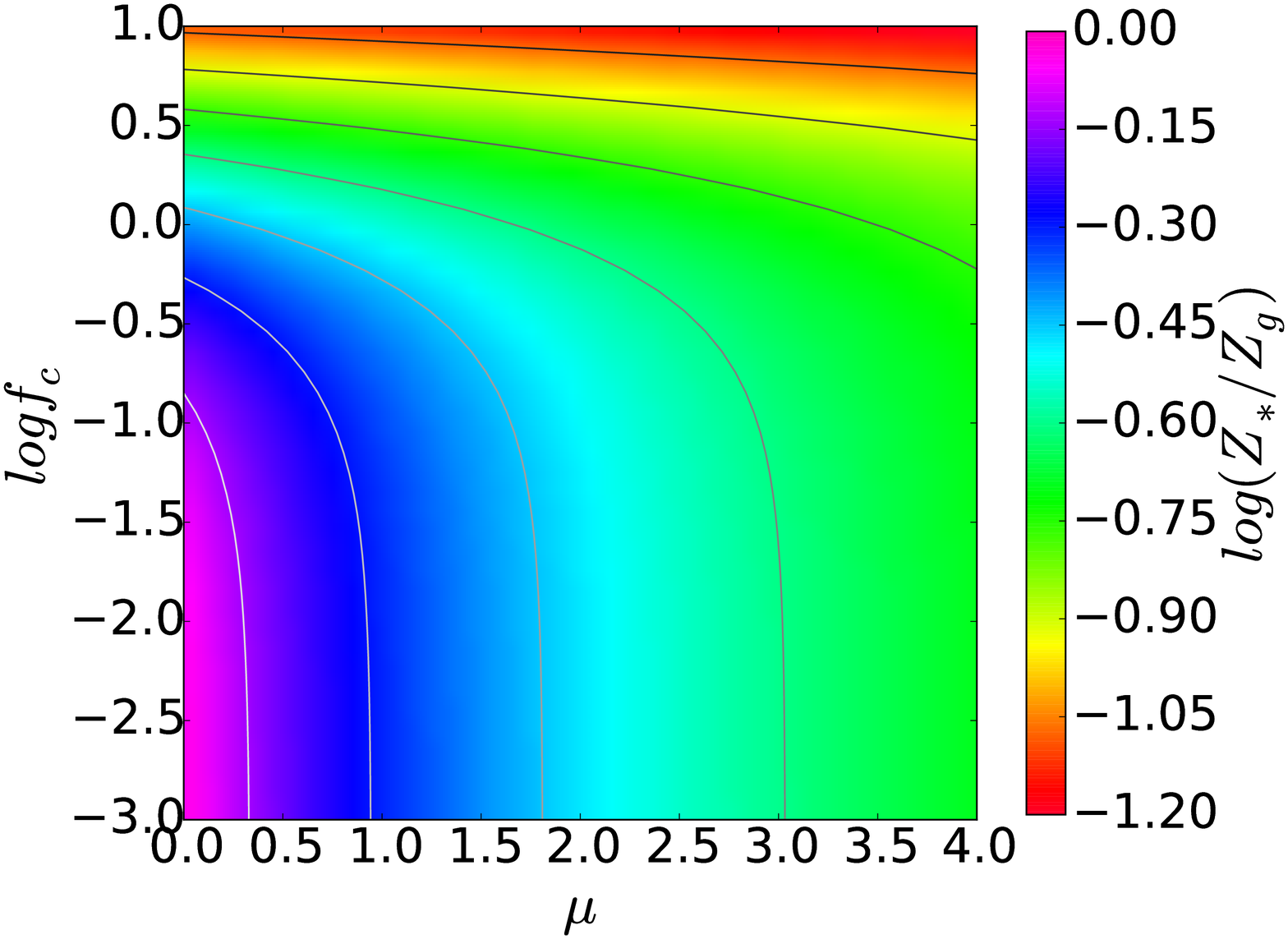}
\caption{The left panel shows the stellar-phase metallicity to gas-phase metallicity ratio 
as a function of stellar mass. 
The local galaxy gas-phase metallicity relations are adopted 
from \citet{Tremonti2004a}, \citet{Maiolino2008a}, \citet{Zahid2013a}, \citet{Andrews2013a}, and \citet{Lee2006a}, 
which are denoted by colored solid lines of T04, M08, Z13, A13, and L06, respectively. 
Either the Gallazzi et al. result (G05) or the Kirby et al. result (K13) for the stellar-phase metallicity is adopted 
for the relevant stellar mass range. 
The right panel shows the stellar-phase metallicity to gas-phase metallicity ratio as a function of 
parameter $\mu$ and $f_{\rm c}$ in Equation (\ref{equ:zratio}). 
 }
\label{fig:fz}
\end{center}
\end{figure*}

The metallicity of a star approximately tells us the metallicity of the ISM at an early time when the star formed, 
while the gas-phase metallicity tells us the current metal content of the ISM. Using these two pieces of 
information, one can gain some insight into the metal enrichment history of galaxies.  
The left panel of Figure \ref{fig:fz} shows the ratio between the stellar-phase metallicity and the gas-phase 
metallicity as a function of stellar mass of local galaxies. 
For the gas-phase metallicity relations, we adopt the results of \citet{Andrews2013a}, \citet{Maiolino2008a},
\citet{Tremonti2004a}, \citet{Zahid2013a}, and \citet{Lee2006a} for local galaxies. 
For the stellar-phase metallicity, we use the \citet{Gallazzi2005a} result and \citet{Kirby2013a} result 
for relevant stellar mass range.  Although these data sets have different 
amplitudes due to different calibrations, they show a systematic trend that low-mass galaxies ($M_*<10^{10}\msun$) 
have a relatively lower stellar-phase metallicity to gas-phase metallicity ratio than the high-mass ones.
This trend indicates that galaxies with different masses follow different ways of metal enrichment.

Assuming that the gas-phase metallicity trajectory in the $Z_{\rm g}-M_*$ plane follows a power-law function, 
we find that the ratio of the stellar phase metallicity and the gas-phase metallicity is 
\begin{equation}
{Z_* \over Z_{\rm g}} = {1 \over \mu +1}\,,
\end{equation} 
where $\mu$ is the power-low index of the trajectory. 
What this equation tells us is that a lower stellar-phase metallicity to gas-phase metallicity ratio 
corresponds to a steeper gas-phase metallicity trajectory (larger $\mu$).  
The left panel of Figure \ref{fig:fz} shows that high-mass galaxies have higher ${Z_* / Z_{\rm g}}$, which 
means they need to have lower $\mu$ than low-mass galaxies. This is consistent with the $\mu-M_*$ relation 
we derived from the gas-phase metallicity relation at different redshifts in \S\ref{sec:result_app2} 
(see the inserted panel of Fig.\,\ref{fig:ztrack}). 
This suggests that high-mass galaxies seem to maintain a slow increase of metallicity for a long time in the past, 
while low-mass galaxies have rapidly enriched their metallicities in recent times.

We can also use a more sophisticated model to describe the trajectory, as the power-law model might be too simple 
to capture the entire evolution, especially at early times. We choose the following model using a power-law to 
describe the late time evolution, and an exponential term to capture the possible rapid increase of metallicity 
following a starburst phase at the early time, i.e.
\begin{equation}
Z_{\rm g}(t) ={Z_{\rm g,0} \over K} \left[ {M_*(t) \over M_{\rm c}} \right]^{\mu} \exp \left[-{M_{\rm c} \over M_*(t) }\right]\,,
\end{equation}
where $K= \left({M_{*,0} \over M_{\rm c}}\right)^{\mu} \exp\left(-{M_{\rm c} \over M_{\rm *,0}}\right)$ is a normalization factor, 
$M_{\rm *,0}$ and $Z_{\rm g,0}$ are the final stellar mass and gas-phase metallicity, 
$\mu$ parameter is the power-law slope of the trajectory of $Z_{\rm g}$ when $M_*>M_{\rm c}$, 
and $M_{\rm c}$ is a characteristic mass scale at which the $Z_{\rm g}$ trajectory switches from an exponential function
to a power-law. Below $M_{\rm c}$, the gas-phase metallicity increases exponentially with 
mass. This model captures the idea that feedback might be powerful enough to 
expel all produced metals at the early epoch of galaxy formation, and it becomes less powerful at late time 
when the galaxy mass is high so that metals can be retained.

Using this model for the trajectory, we find that 
\begin{equation}\label{equ:zratio}
{Z_* \over Z_{\rm g}} =  f_{\rm c} \exp(f_{\rm c}) \Gamma (-\mu-1, f_{\rm c})\,,
\end{equation}
where $f_{\rm c}\equiv {M_{\rm c} / M_{*,0}}$, and $\Gamma(s,x)$ is the upper incomplete gamma function: 
$\Gamma(s,x)=\int_x^{\infty} t^{s-1} e^{-t}{\rm d}t$. 
We show the two-dimensional function in the right panel of Figure \ref{fig:fz}. 
The contours in the figure show the combinations of $f_{\rm c}$ and $\mu$ that give rise to a constant ${Z_* / Z_{\rm g}}$. 
As one can see, when ${Z_* / Z_{\rm g}}\sim 1$, the gas-phase metallicity track is required to be flat ($\mu\sim0$ 
and $f_{\rm c}$ is very small). 
When ${Z_* / Z_{\rm g}}$ becomes smaller, the gas-phase metallicity trajectory has to be a steep function of 
stellar mass, either with a large power-law slope $\mu$ or a higher $f_{\rm c}$. To result in this type of metallicity 
trajectory, a large fraction of stellar mass needs to form in an early starburst with very low metallicity, and 
to retain metals in the ISM to rapidly enrich the gas-phase metallicity at late time.

This inference has two implications. First, the low $Z_*/Z_{\rm g}$ ratio of low-mass galaxies seems to 
indicate that a relatively large fraction of stellar mass of these galaxies formed in early starbursts when the ISM 
metallicity is still low. 
Second, low-mass galaxies, as opposed to high-mass galaxies, are more rapidly enriching their metallicity, 
indicating a downsizing effect that low-mass galaxies are less evolved than the high-mass ones also in metallicity.

\section{Discussion}\label{sec:discussion}

In this paper, we have introduced a simple analytic model to follow the metallicities of galaxies. 
When the model is matched to the gas-phase and stellar-phase metallicities as 
functions of stellar mass of galaxies at different redshifts, these observational data 
in turn can constrain star formation and feedback.

With minimum assumptions about how metals produced by star formation are mixed into the ISM 
and how outflow can affect the baryonic content of galaxies,  
we demonstrate that metallicities of galaxies provide useful constraints on the strength of galaxy outflows. 
Using the model, we have derived two approaches to constrain the outflow mass-loading factor 
as a function of galaxy stellar mass. 
In the first approach, we find the stellar-phase metallicity, gas-phase metallicity and the cold-gas-to-stellar 
mass ratio as functions of galaxy stellar mass at a given redshift can already provide strong constraints 
on the upper limit for the mass-loading factor. 
The fundamental reason that the combination of these data
can constrain outflow is because the gas-phase metallicity represents the metal content of galaxies 
at a fixed time, and the stellar-phase metallicity represent a time-averaged value for the metal content 
of galaxies. The combination of these two quantities provide constraints on how the metal content changes 
over time when galaxies build up their stellar mass. The metal mass that is not found in galaxies should 
have been carried away by SN ejecta or galactic outflow. 
The second approach relies on the gas-phase metallicity relation at different 
redshifts. Once the manner in which galaxies change their metallicity as they increase their stellar mass is determined, 
the metallicity relations at different epochs can also constrain the strength of outflow. 

Combining multiple observational measurements for the stellar mass-metallicity relations, 
we have derived the outflow mass-loading factor based on the simple model. 
The results show that, in spite of the large uncertainties in the metallicity measurements, 
the inferred outflow mass-loading factors have a well defined upper limit. Even with generous (high) 
metal yield and assuming all produced metals are mixed into the ISM, 
the inferred mass-loading factor cannot be higher than 20 for $10^9\msun$ galaxies 
and it drops quickly for higher galaxy masses. 
On the very low mass end, where re-ionization is effective to block a significant fraction of baryons 
from collapsing into the galaxies, strong outflow is not a plausible explanation for the cold baryon 
mass fraction and the metallicity relations. To explain the data, we find that a significant fraction 
of produced metals need to be ejected directly out of the galaxies with SN ejecta without mixing 
in the ISM, which is consistent with the results of \citet{Dalcanton2007a}, who found that 
metal-enriched outflow is the only viable mechanism to reproduce the low retained yields of gas-rich 
low-mass galaxies. Observations of metal content of winds from dwarf starburst also suggest that 
almost all the metals produced in starbursts are ejected directly out of the galaxies \citep[e.g.][]{Martin2002a}, 
providing observational support of metal-enriched ejection. 
If a moderate fraction (half) of metal mass is involved in the metal-enriched ejection, 
the mass-loading factor is expected to be at most only a few ($\eta<8$). 
This low mass-loading factor conflicts with many galaxy formation models, which require 
much higher mass-loading factors, especially for low-mass galaxies ($M_*\leq10^{10}\msun$), 
to explain low-mass end of the stellar mass function \citep{Lu2014a, Benson2014a}.
In many successful galaxy formation models, strong outflow is needed because baryons other than 
observed stars and cold gas cannot stay in the halos without forming stars. The only way to keep 
low-mass halos from having a too high baryon mass fraction is to eject a large amount of baryonic mass out of the halo. 
What the metallicity relations seem to show in this paper, however, is that when galaxies increase their 
stellar mass (and simultaneously produce metals), the metal mass in galaxies increases 
substantially. The increase in the metal mass suggests that there should not be very strong outflow to take away metals.

The data show that low-mass galaxies tend to have a lower stellar-phase to gas-phase metallicity ratio. 
Using simple analytic models, we conclude that this trend suggests 
that low-mass galaxies tend to increase their gas-phase metallicity rapidly in late times, 
while high-mass galaxies have gone through the rapid metal enrichment phase in early times, 
which indicates an imprint of the downsizing effect in the metallicity evolution of galaxies. 
The analysis also suggests that before the rapid enrichment epoch, a fraction of stellar mass is 
formed at early times when the gas-phase metallicity was low. 
This is consistent with recent observational results that 
local dwarf galaxies have more than half of their stellar mass form in early phase prior to $z=2$ \citep{Weisz2014a}.

 Finally, we stress the importance of the accuracy of the metallicity relations and 
cold gas mass fraction in observational measurements. 
We have adopted simplified version of existing observational results to demonstrate 
the constraining power of the data. Better observational determination of these scaling relations will definitely 
improve the constraints when more sophisticated model inferences, such as Bayesian model inference \citep[e.g.][]{Lu2011b}, are adopted.

\section*{Acknowledgements}
We thank Andrew McWilliam, Houjun Mo, Josh Simon for useful discussions.  
\bibliography{references}


\end{document}